\documentclass[runningheads]{svmult}

\usepackage{makeidx}   
\usepackage{graphicx}  
\usepackage{subeqnar}  
\usepackage{multicol}  
\usepackage{physprbb}  
\makeindex             

\begin{document}
\title*{Doppler Tomography of Dwarf Nova IY UMa during Quiescence}
%
%
\toctitle{Doppler Tomography of Dwarf Nova IY UMa during Quiescence}
%
%
\titlerunning{Doppler Tomography of Dwarf Nova IY UMa during Quiescence}
%
\author{Daniel J. Rolfe\inst{1}
\and Timothy M. C. Abbott\inst{2}
\and Carole A. Haswell\inst{1}}
\authorrunning{Daniel J. Rolfe et al.}
%
%
\institute{Department of Physics and Astronomy, The Open University, Walton Hall,\\Milton Keynes, MK7 6AA, UK
\and Nordic Optical Telescope, Roque de Los Muchachos \& Santa Cruz de La Palma,\\Canary Islands, Spain}

\maketitle              

\begin{abstract}
  Quiescent Doppler tomography of the newly discovered
  deeply-eclipsing SU~UMa system IY~UMa reveals properties of the
  region where the accretion stream from the donor impacts the edge of
  the disc. A very strong bright spot is produced and the Keplerian
  disc emission in the impact region is disrupted or obscured. The
  differing properties of H$\alpha$, H$\beta$ and He I emission will
  allow physical parameters of the converging flow region to
  be studied.
\end{abstract}

\section{Introduction}

\begin{figure}[b]
\begin{center}
\includegraphics[width=0.7\textwidth]{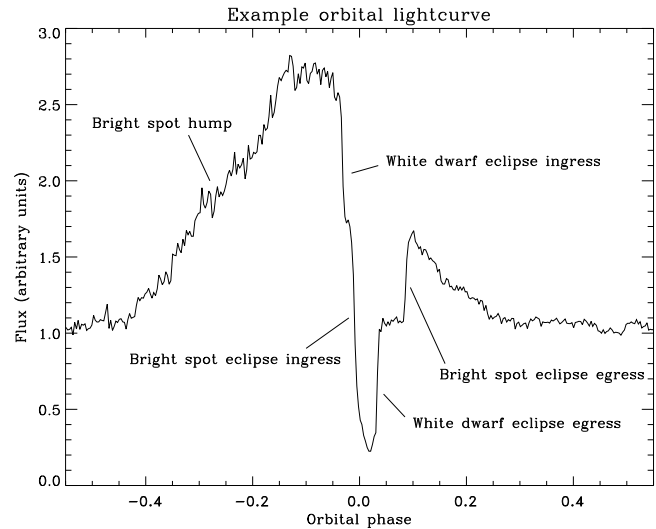}
\end{center}
\caption[]{Example orbital lightcurve of IY~UMa close to quiescence. Note the strong orbital hump as the stream-disc impact region comes into view, the eclipse of this region and the eclipse of the white dwarf. Taken from \cite{rhp}.}
\label{orbcurve}
\end{figure}

IY~UMa was observed for the first time with a superoutburst in January
2000 identifying it as a SU UMa type dwarf nova cataclysmic
variable~\cite{uemura} with orbital period 1.77 hours. It exhibits
deep eclipses, with the eclipse of the white dwarf and the stream-disc
impact region being clearly identifiable in quiescence
(Fig.~\ref{orbcurve}); this behaviour is similar to that of the other
eclipsing SU~UMa systems OY~Car~\cite{schoembs} and
Z~Cha~\cite{wood1}.

The orbital parameters have been estimated~\cite{patt} as
$M_{wd}=0.93\pm0.14 M_\odot$, $M_{donor}=0.12\pm0.03 M_\odot$ and
inclination $i=87^{\circ} \pm 3^{\circ}$.

\section{Spectroscopy}
\subsection{Observations}
We obtained 3 orbits of coverage on 19th March 2000 using ALFOSC on
the Nordic Optical Telescope in La Palma. There are 80 spectra with
resolution $\sim$4\AA~covering wavelength range 3900\AA~to 6850\AA.

\begin{figure}[b]
\begin{center}
\includegraphics[width=.7\textwidth,angle=90]{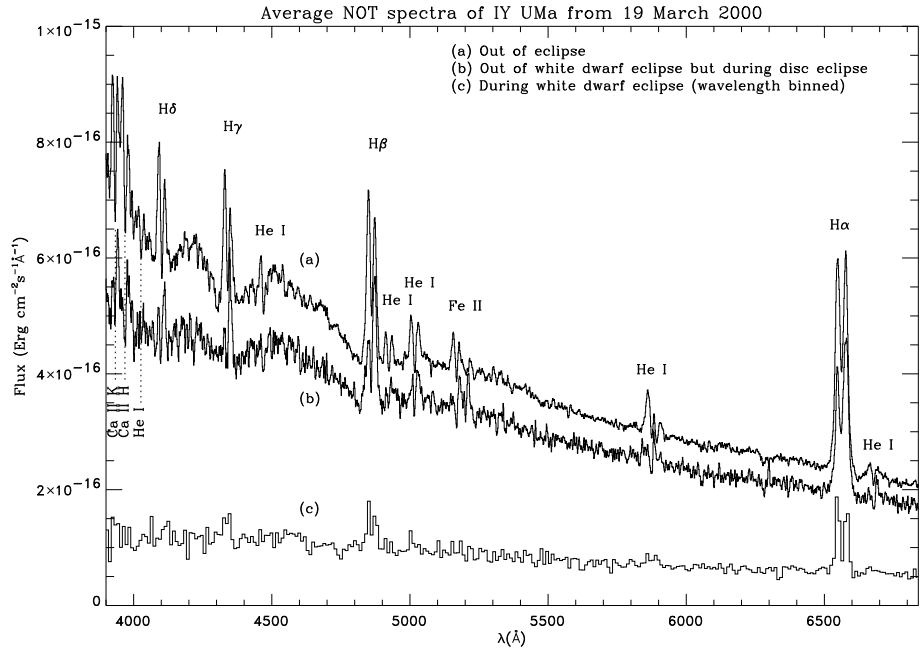}
\end{center}
\caption[]{Average spectra of IY UMa on 19th March 2000.}
\label{spectra}
\end{figure}

\subsection{Average Spectra}
Figure~\ref{spectra} shows average spectra covering different sections
of the orbit. The top spectrum (a) is out of eclipse data; the middle
spectrum includes phases where the disc is eclipsed but the white
dwarf is {\bf not}; (c) is the average during white dwarf eclipse.
The spectra show double-peaked Balmer, He I and Fe II emission, the
double-peaked structure signalling that this emission comes from the
accretion disc.  There are also strong broad Balmer absorption wings
in H$\beta$, H$\gamma$ and H$\delta$ in spectra (a) and (b) which only
disappear during white dwarf eclipse (c), telling us that these
features come from close to the white dwarf, if not from the white
dwarf itself.  There is also deep core absorption clearly seen in
H$\delta$, He I and Fe II.  These features are almost identical to
those seen in OY~Car~\cite{hessman}. Z~Cha also shows very similar
features in quiescence~\cite{mhs}.

\subsection{Systemic Velocity}

The radial velocity of the H$\alpha$ line was measured for each
spectrum by fitting a double gaussian profile and using the velocity
of the midpoint.  A sinusoid of the form
$V=V_\mathrm{0}-V_\mathrm{1}sin 2\pi(\phi-\phi_\mathrm{0})$ was fitted
to those velocities outside eclipse.  We obtain systemic velocity
$V_\mathrm{0}=15.8\pm1.3$ km s$^{-1}$, velocity amplitude
$V_\mathrm{1}=100.4\pm1.4$ km s$^{-1}$ and
$\phi_\mathrm{0}=0.119\pm0.003$. The large phase shift,
$\phi_\mathrm{0}$ relative to white dwarf mid-eclipse, tells us that
the emission does not follow the motion of the white dwarf and so
$V_\mathrm{1}$ cannot be treated as the white dwarf velocity. This
phase shift is the same as that seen in the unusual dwarf nova
WZ~Sge~\cite{skid2} and also similar to those in SU~UMa systems OY~Car
(measurements summarized in \cite{hessman}) and Z~Cha~\cite{mhs}.
Similar phase shifts are seen in quiescent low mass X-ray transients
e.g. V616 Mon~\cite{haswell}.

\subsection{Trailed Spectra}

\begin{figure}[b]
\begin{center}
\includegraphics[width=1.0\textwidth]{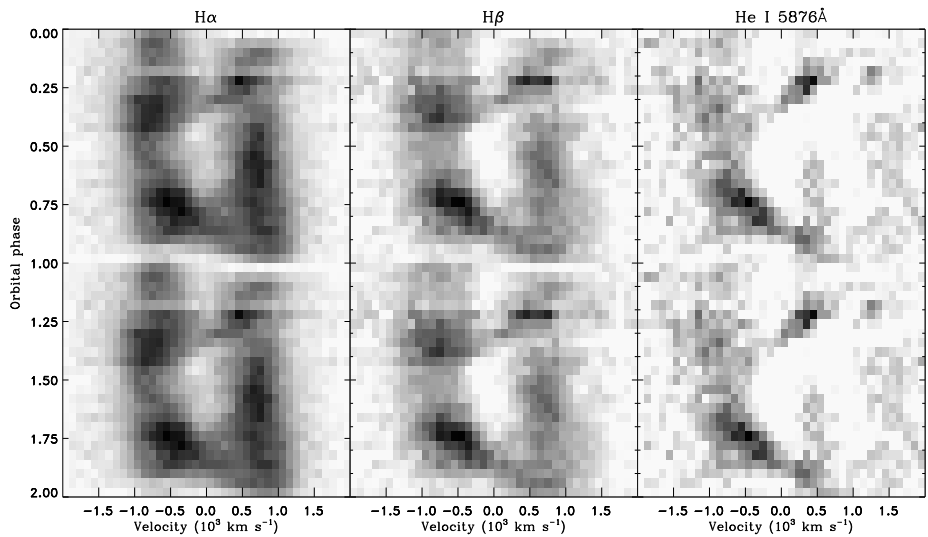}
\end{center}
\caption[]{Phase-folded, velocity-binned and continuum-subtracted trailed spectra. Black corresponds to the maximum flux, while white is the continuum level. Absorption (values below continuum) are displayed in white.}
\label{trails}
\end{figure}

In Fig.~\ref{trails} we show trailed spectra of H$\alpha$, H$\beta$
and He~I~5876\AA. In all three lines we see the eclipse beginning in
the blue and ending in the red as first the side of the disc coming
towards us and then the side of the disc moving away is occulted by
the donor star. In H$\alpha$ we clearly see the double-peaked emission
component from the disc, with average peak-peak separation in the
phase range 0.4--0.6 (where the bright spot is on the far side of the
disc and so will have minimal effect on the measurement) of 1440
km~s$^{-1}$ corresponding to a radius of $0.45a$ assuming a Keplerian
velocity field ($a$ is the orbital separation).  There is also a
strong S-wave component corresponding to a localized region of
emission. The S-wave is weakest around orbital phase 0.5.
The structure in H$\beta$ is very similar, except that the disc
emission is fainter compared to the S-wave.
He I shows very little evidence of disc emission, but again exhibits a
strong S-wave which is weakest around phase 0.5. Without the
complication of the disc component, we see that the brightness of the
S-wave closely follows the orbital hump in the continuum lightcurve.
The He I emission reveals that the eclipse of the S-wave is late,
placing it in the correct region of the disc to be the stream-disc
impact region.  The strong low velocity absorption is present.

\subsection{Doppler Maps}

\begin{figure}[b]
\begin{center}
\includegraphics[width=1.0\textwidth]{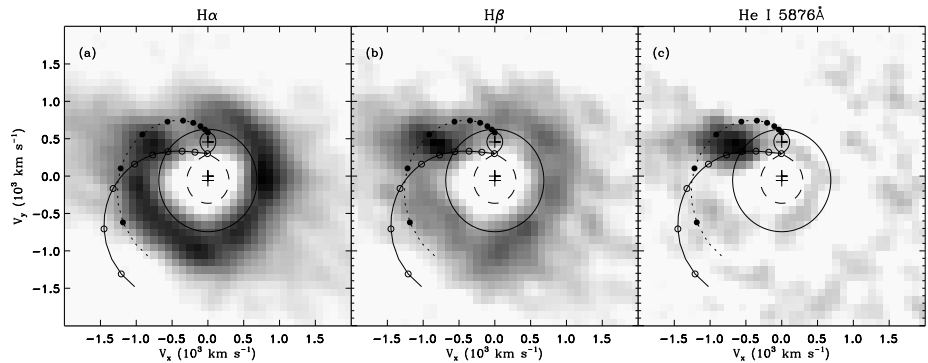}
\end{center}
\caption[]{Fourier-filtered back-projections. Black corresponds to the maximum flux in each map, while white is the continuum level. Absorption (values below continuum) are displayed in white. The small teardrop shape shows the velocity of the donor star, the arc with unfilled circles shows the velocity along the stream trajectory with circles denoting steps of $0.1a$ along the stream. The arc with solid circles is the Keplerian velocity along the stream trajectory. The circle is the Keplerian velocity at the tidal truncation radius.}
\label{backpros}
\end{figure}

We used the Fourier-filtered back-projection method~\cite{marshhorne}
to obtain maps of the velocity-space distribution of emission in each
line, shown in Fig.~\ref{backpros}. Only out-of-eclipse spectra were
used to generate the Doppler maps, since occultation by the donor star
violates a core assumption of the Doppler tomography method, which
assumes that all regions have the same visibility at all orbital
phases.

\begin{figure}[b]
\begin{center}
\includegraphics[width=0.6\textwidth]{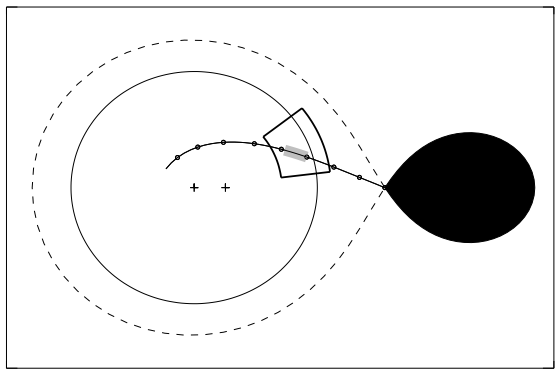}
\end{center}
\caption[]{Spatial geometry interpreted from Doppler map. Stream from donor star (with circles at separation $0.1a$ marked along it) first impacts disc at hot spot (grey region). Circle marks radius at which Keplerian velocity is equal to half the H$\alpha$ peak separation.}
\label{geometry}
\end{figure}

The H$\alpha$ map (Fig.~\ref{backpros}a) shows a ring of disc emission
with Keplerian velocities corresponding to locations within the tidal
radius. The stream-disc impact should be seen between the two arcs
corresponding to the stream trajectory and its Keplerian velocity in
the top left of the map, but there is no significantly brighter region
here in the H$\alpha$ map. The disc emission just to the left of the
donor star is very much weaker than elsewhere. Assuming a Keplerian
velocity field, this corresponds to weaker emission from the region of
the disc marked with a thick black outline in Fig.~\ref{geometry}, and
is coincident with the stream-disc impact. We do not expect a
Keplerian velocity field where the stream and disc merge, and any
underlying Keplerian emission could be obscured by an optically thick
region around the impact. It is therefore no surprise that we see
weaker disc emission in this velocity region. At low velocities (less
than $\sim$500 km~s$^{-1}$) we see strong absorption.  This absorption
is of similar strength to the disc emission.

The H$\beta$ map (Fig.~\ref{backpros}b) again shows the Keplerian
emission from the disc and the strong low-velocity absorption. Most
notable, however, is the bright spot located exactly where we expect
to see the stream-disc impact. Combined with spatial information
provided by the late eclipse of the S-wave which corresponds to this
hot spot, and the fact that the variation in brightness of the S-wave
also closely follows the orbital hump, we conclude that this emission
is coming from stream-disc impact. The position along the stream
trajectory at which the H$\beta$ bright spot is brightest corresponds
to the grey region in Fig.~\ref{geometry}, just within the disc radius
deduced from the peak-peak separation of the H$\alpha$ disc emission.

The He I 5876\AA~map (Fig.~\ref{backpros}c) shows no disc emission at
all. It has the strong low-velocity absorption and also the bright
spot due to the stream-disc impact.

The behaviour of the H$\alpha$ and H$\beta$ maps is the same as that
seen in WZ~Sge~\cite{skid1} and also very similar to that in
HE~1047~\cite{skid3}.  The strong disc component in both lines, and
the increase in the relative strength of the stream-disc impact to the
disc in H$\beta$ is seen in both of the systems, while the weaker
region of disc emission in H$\alpha$ between the hot spot and donor is
seen in WZ~Sge but not in HE~1047.

\subsection{Conclusions}
The deeply-eclipsing dwarf nova IY UMa has average optical spectra
during quiescence similar to those of the two similarly high
inclination dwarf novae OY Car and Z Cha. Time-resolved spectroscopy
and the method of Doppler tomography reveals `classic' accretion flow
behaviour in this system. The stream from the donor star impacts the
edge of the Keplerian accretion disc, dissipating its energy in a
fairly concentrated region (the bright spot seen in
Figs.~\ref{backpros}b and \ref{backpros}c). This stream-disc impact
disrupts and/or obscures the Keplerian emission leading to the weak
disc emission in the top left of Fig.~\ref{backpros}a.

\subsection{Acknowledgements}
The data presented here have been taken using ALFOSC, which is owned
by the Instituto de Astrofisica de Andalucia (IAA) and operated at the
Nordic Optical Telescope under agreement between IAA and the NBIfA of
the Astronomical Observatory of Copenhagen. The Nordic Optical
Telescope is operated on the island of La Palma jointly by Denmark,
Finland, Iceland, Norway, and Sweden, in the Spanish Observatorio del
Roque de los Muchachos of the Instituto de Astrofisica de Canarias.
The authors thank Joe Patterson and Jonathan Kemp for the data used to
produce Fig.~\ref{orbcurve}.  The Doppler maps were produced using the
software package MOLLY by Tom Marsh. We acknowledge the data analysis
facilities at the Open University provided by the OU research
committee and the OU computer support provided by Chris Wigglesworth.
DJR is supported by a PPARC studentship. CAH acknowledges support from
the Leverhulme Trust F/00-180/A.

%

\end{document}